\documentclass[aps,pra,twocolumn,superscriptaddress,showpacs,amsmath,amssymb,amsfonts,longbibliography]{revtex4-1} 

\usepackage[utf8]{inputenc}
\usepackage{amsthm}
\usepackage{rotating}
\usepackage{graphicx} 

\usepackage{times}
\usepackage{mathptmx}

\usepackage[colorlinks=true,citecolor=blue,linkcolor=blue,urlcolor=blue]{hyperref}%
\usepackage[T1]{polski} 
\usepackage[english]{babel}
\usepackage{bbold}  
\usepackage{pgfplots}

\usepackage[normalem]{ulem} 
\usepackage{natbib}  
\usepackage{cancel} 
\newcommand{\bra}[1]{\left\langle#1\right|}
\newcommand{\ket}[1]{\left|#1\right\rangle}
\begin{document}
\title{Disorder-assisted graph coloring on quantum annealers}
\author{Andrzej Wi\k{e}ckowski}%
\affiliation{
Department of Theoretical Physics, 
Faculty of Fundamental Problems of Technology,
Wrocław University of Science and Technology,  
50-370 Wrocław,  Poland}
\author{Sebastian Deffner}
\affiliation{Department of Physics, University of Maryland, Baltimore County, Baltimore, MD 21250, USA}
\author{Bartłomiej Gardas}
\affiliation{
Institute of  Physics,  University  of  Silesia,
40-007  Katowice,  Poland}
\affiliation{Jagiellonian University,  Marian Smoluchowski Institute of Physics, \L{}ojasiewicza 11, 30-348 Krak\'ow, Poland}

\begin{abstract}
We are at the verge of a new era, which will be dominated by Noisy Intermediate-Scale Quantum Devices. Prototypical examples for these new technologies are present-day quantum annealers. In the present work, we investigate to what extent static disorder generated by an external source of noise does not have to be detrimental, but can actually assist quantum annealers in achieving better performance. In particular, we analyze the graph coloring problem that can be solved on a sparse topology (i.e. chimera graph) via suitable embedding. We show that specifically tailored disorder can enhance the fidelity of the annealing process and thus increase the overall performance of the annealer.
\end{abstract}

\maketitle

\section{Introduction}
The first concept of quantum computing was formulated several decades ago in an attempt to faithfully simulate many-body quantum systems, which is known to be an impossible feat with classical computers~\cite{Feynman60,Feynman1982}. However, only very recently novel technologies have become available that promise to make quantum computers a practical reality~\cite{Sanders2017}.
Quite remarkably, already the first generation of fully operational quantum computers is expected to outperform (for specific tasks) even the most advanced, state-of-the-art classical computers~\cite{Sanders2017,diventra18}.
To be ready for the first physical realizations of such powerful information technology, quantum computer science has been developing a plethora of quantum algorithms for a wide variety of optimization problems~\cite{Mosca2008}.
Famous examples include the Deutsch--Jozsa algorithm~\cite{deutsch1992rapid} to evaluate a function, the Grover algorithm~\cite{grover1996fast} for searches of a (possibly large) database, or Shor’s algorithm~\cite{shor1999polynomial} designed for prime factorization.

In the present work we will focus on adiabatic quantum computation (AQC)~\cite{farhi2000quantum}, which relies on quantum annealing \cite{Kadowaki1998}. In comparison to other computational paradigms, AQC is technologically slightly more advanced due to the commercial availability of D-Wave's quantum annealers~\cite{PhysRevB.82.024511,johnson2011quantum,boixo2013experimental}.
Adiabatic quantum computing is a computational paradigm~\cite{RevModPhys.90.015002} that has the potential to solve many problems that a universal quantum computer can also solve~\cite{aharonov2008adiabatic}. Although, a polynomial time penalty may be necessary to achieve this, with AQC one can still outperform classical computers in many practical cases~\cite{Farhi472}.

AQC relies on the quantum adiabatic theorem~\cite{farhi2000quantum}. In this paradigm,
a quantum system is prepared in the ground state of an initial (``easy'') Hamiltonian $H_{\rm i}$. Then, 
the system is let to evolve adiabatically---infinitely slowly---towards the ground state of the final Hamiltonian $H_{\rm f}$. The latter system encodes the problem of interest and its ground state stores the desired solution (i.e. an answer to the problem). Devices that can realize such evolution are called quantum annealers~\cite{Kadowaki1998}. Quantum annealers are typically designed with one and only one particular task in mind---namely, to solve combinatorial optimization problems from the {\bf NP} complexity class~\cite{Barahona1982,PhysRevA.78.012352}. These problems are ``very hard'' to solve with classical computers, however their solutions can still be verified (in polynomial time). 
 
Several advantages of quantum annealing over other computational paradigms have been identified~\cite{gardas2018quantum1,PhysRevX.8.031016,Ronnow420,PhysRevLett.106.050502}. However, currently available technology still exhibits hardware issues, of which the most important one is static disorder~\cite{brugger2018quantum,kudo2018constrained,kudo2019graph,gardas2018defects,gardas2018quantum2}. Rather counter-intuitively, however, it also has been shown that static disorder is not always detrimental, but can rather be a valuable resource in achieving quantum tasks~\cite{Novo2016,Almeida2017}.

In the present work, we study the influence of static disorder on the annealing dynamics and analyze its effect on the performance of near-term quantum annealers. To this end we mainly focus on a selected problem of graph coloring~\cite{graphcoloring}. This a fundamental problem in modern computer science with various applications in many different areas, e.g. in scheduling~\cite{marx2004graph}, pattern~\cite{BRASSILVA2006776} and frequency~\cite{TaehoonPark1996KJ00001203022} matching, or memory allocation~\cite{Chaitin:1982:RAS:872726.806984}, to name just a few.

The main objective of the graph coloring problem is to find a minimal number of colors, {\it chromatic number} -- $\chi(G)$, that are required to color a graph $G$, so that no adjacent sites share the same color. In this context, colors can encode any arbitrary information. Typical examples are shown Fig.~\ref{fig:graph-example}. Remarkably, we will find that for the graph coloring problem D-Wave like annealers may actually be robust against certain type of noise. Even more importantly, we will see that particular types of disorder can assist the adiabatic computation to achieve better performance.

\section{Disorder graph coloring problem \label{ch:coloring}}

\begin{figure}[t!]
    \centering
    \includegraphics[width=\linewidth]{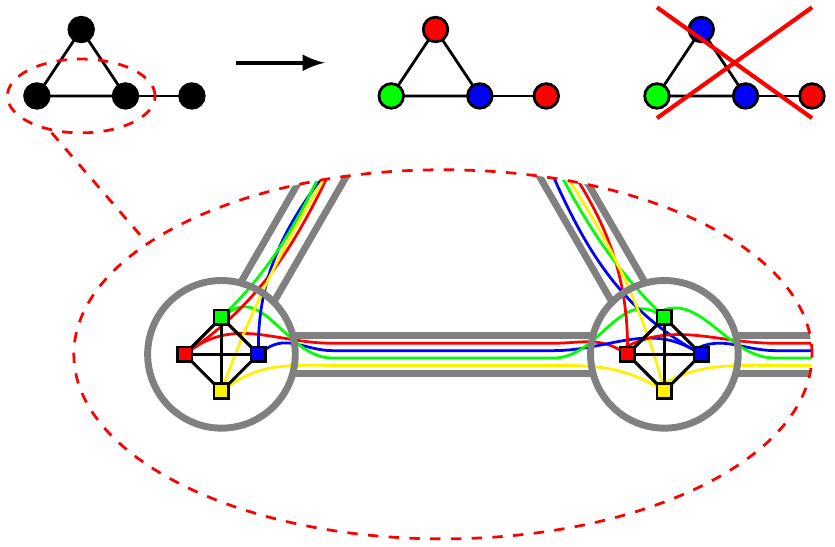}
    \caption{
    Graph coloring problem exemplified with $N=4$ vertices and $K=3$ colors. A different color is assigned to adjacent vertices. Other configurations are {\it not} valid solutions.
    A binary variable $X_{ic}=1$ represents a vertex $i\le N$ having a color $c\in\{1,2,\dots,K\}$. Otherwise we set $X_{ic}=0$, cf. Eq.~(\ref{eq:qubocolor}). 
    }
    \label{fig:graph-example}
\end{figure}

The dynamics of quantum annealers is typically described by the following Hamiltonian,
\begin{equation}
\label{eq:AQC}    
\hat H(s) = f(s) \hat H_{\text i} + [1-f(s)] \hat H_{\text f},
\quad 
s\in[-1,1],
\end{equation}
where $f(s)\in[0,1]$ could be an arbitrary function such that 
$f(-1)=1$ and $f(1) = 0$~\cite{lanting2014entanglement}. 
Typically, $f(s)=s+1$ where $s(t)=t/\tau$ and $\tau$ is the annealing time~\cite{gardas2018defects}. For the present purposes, initial and final Hamiltonian are instances of the Ising spin-glass~\cite{PhysRevE.58.5355}, where in particular, 
\begin{equation}
\label{eq:hamisingchain}
\hat H_{\text f} = \sum_{\langle i,j\rangle\in\mathcal{E}}J_{ij}S_i^z S_j^z+\sum_{i\in\mathcal{V}}h_iS_i^z,
\quad
\hat H_{\text i}  = 4 \sum_{i\in\mathcal{V}} S_i^x, 
\end{equation}
\noindent
Here, the problem Hamiltonian, $\hat H_{\text f}$, is defined on a graph, $\mathcal{G}=(\mathcal{E}, \mathcal{V})$, specified by its edges, $\mathcal{E}$, and vertices, $\mathcal{V}$. This simple model can already be realized with present-day quantum annealers~\cite{PhysRevX.8.031016}, where the graph $\mathcal{G}$ is set to reflect the chimera~\cite{Choi2008,Choi2011} or pegasus topology~\cite{dattani2019pegasus}. The programmable input parameters~\cite{PhysRevB.81.134510} are the elements of the coupling matrix, $J_{ij}$, and the onsite magnetic fields, $h_i$. 
Spin operators are denoted by $S_i^z$, $S_i^x$ and they describe spins in the $z, x$ directions respectively.

All Ising variables can admit only two values ($s_i=\pm 1$). Since there are, however, typically more than two colors necessary to solve a graph coloring problem, one cannot map it \emph{directly} onto the Ising Hamiltonian. Thus, graph coloring problems are first expressed as spin-lattices, where the spins can take more than two values. These so-called  Potts models~\cite{npising,RevModPhys.54.235}
can then be mapped onto the Ising Hamiltonian using a suitable embedding (i.e. with the help of auxiliary variables).

\begin{figure}[t]
\centering
\includegraphics[width=\columnwidth]{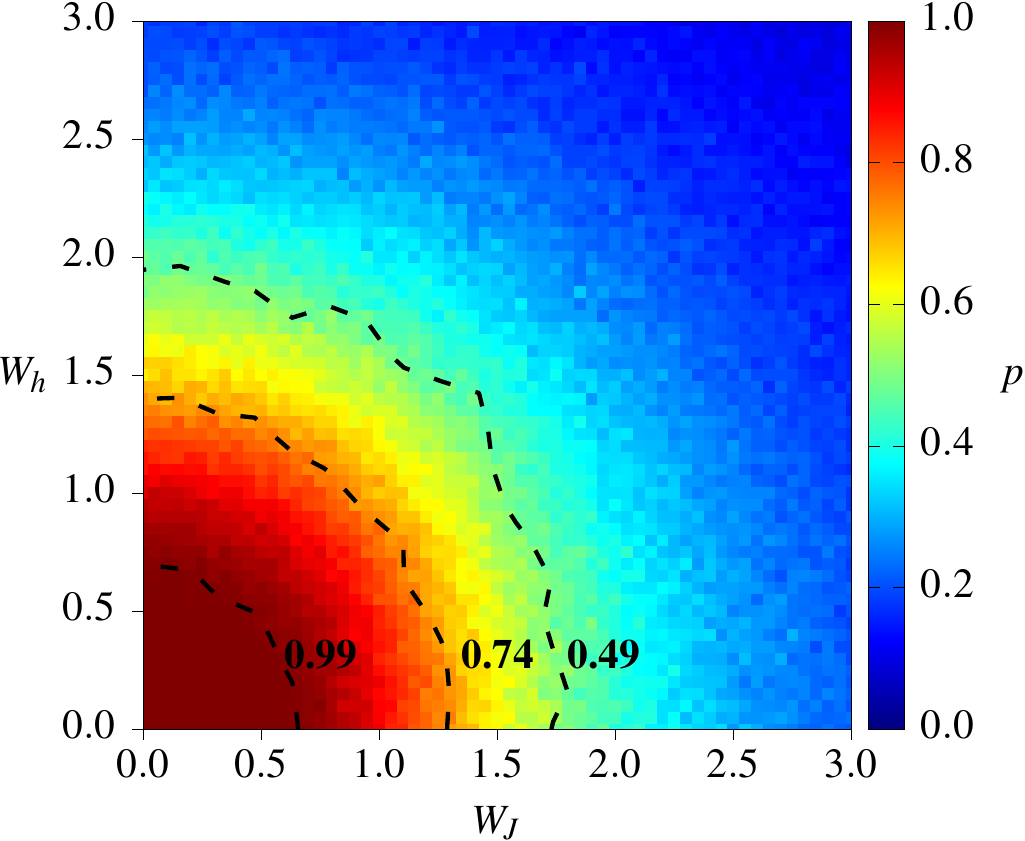}
\caption{
\label{fig:cmap}
Influence of the disorder's amplitudes $(W_h,W_J)$ on the ground state
properties of the system's Hamiltonain~(\ref{eq:hamisingchain}).
Here, $p$ is the probability for the ground state of the disordered system being any of the solutions to the disorder-free problem [cf. Eq.~(\ref{eq:disorder})]. Contours show probabilities $0.99$, $0.74$ and $0.49$ respectively. The result has been obtained for the triangle topology.
}
\label{fig:gsdis}
 \end{figure}

When designing quantum algorithms, it is often convenient to work with the Quadratic Unconstrained Binary Optimization framework or QUBO~\cite{WANG20093746}.
Here, we introduce a binary variable $X_{ic}=1$ if a vertex $i\in\{1,2,\dots N\}$ is colored with a color $c\in\{1,2,\dots,K\}$ and we set $X_{ic}=0$ otherwise. Then
the graph coloring problem can be formulated in the following simple terms (cf. Fig.~\ref{fig:graph-example})
\begin{equation}
\label{eq:qubocolor}
\hat H_{\rm f}^{\text Q} =
\sum_{i=1}^N\left(1-\sum_{c=1}^K X_{ic}  \right)^2
+ 
\sum_{\langle i,j\rangle} \sum_{c=1}^K X_{ic} X_{jc},
\end{equation}
\noindent
where $\langle i, j \rangle$ indicates summation over 
all connected vertices. If the ground state of the Hamiltonian in Eq.~(\ref{eq:qubocolor}), corresponding to the energy $E=0$, exists then the graph $G$ can be properly colored with at least $K$ colors. 
The purpose of the first term in the above Hamiltonian is to assure that each vertex $i$ is colored with only one specific color $c$, as only then $\sum_{c=1}^K X_{ic}=1$. The second term introduces an energy penalty whenever neighboring vertices have the same color $c$. Similar encoding strategies have also been discussed in the context of quantum error correcting codes for quantum annealers~\cite{Pudenz2014}.

\begin{figure*}[t]
    \centering
    \includegraphics[width=0.95\textwidth]{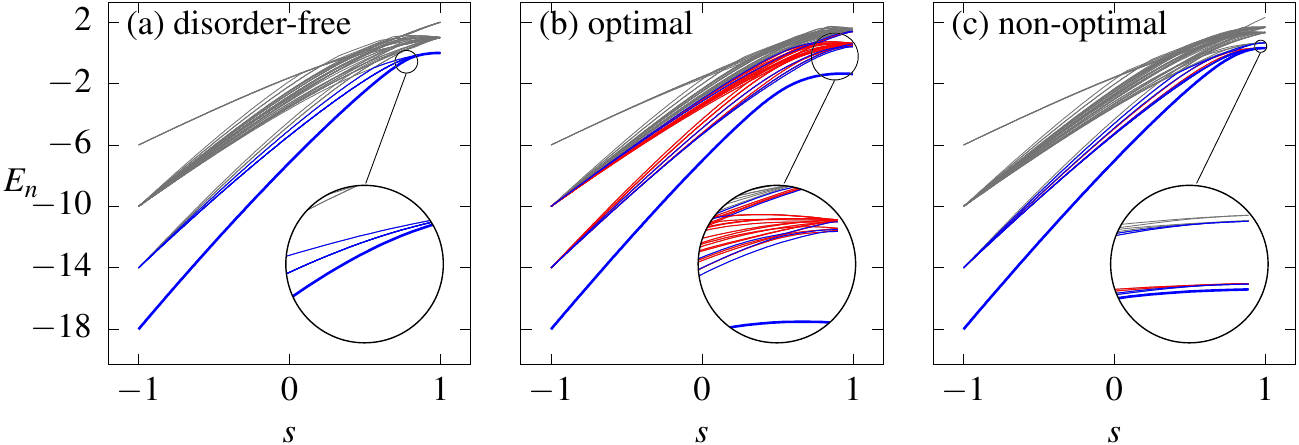}
    \caption{Structure of the low energy spectrum for the disorder-free, (a), and disordered Hamiltonians [(b), (c)], cf.~Eq.\ref{eq:AQC}. The ground state, $E_n(s=1)$, is degenerate for the disorder-free case and thus encodes all different solutions (marked here as blue) to the graph coloring problem. The degeneracy is then removed when disorder is incorporated into the system, cf. Eq.~(\ref{eq:dism2}). Optimally, both the ground state and also excited states encode correct solutions with no ``impurities'' in between (i.e. low energy states representing \emph{incorrect} solutions -- marked as red). This situation increases the effectively gap -- $\Delta$ (defined as the difference between the ground state and the first accessible state) decreasing the computational/annealing time -- $\tau$, cf. Fig.~\ref{fig:all}. In contrast, non-optimal realizations results in impurities causing the effective gap to shrink. This leads to an increase of the annealing time $\tau$. All plots has been obtained for the triangle topology.
    }
    \label{fig:En}
\end{figure*}
 
Having formulated the graph coloring problem in terms of binary variables, one can convert it back into the Ising Hamiltonian, which is more common for quantum annealers. Namely,  
\begin{equation}
\label{eq:hamcolorspin}
\begin{split}
\hat H_{\rm f}^{\text I} 
    &=
    \sum_{i=1}^N J_{ii}\sum_{c_1< c_2} S_{ic_1}^zS_{ic_2}^z 
    +
    \sum_{\langle i,j\rangle} J_{ij}\sum_{c=1}^K S_{ic}^zS_{jc}^z \\
    &+
    \sum_{i=1}^N{h_i} \sum_{c=1}^K S_{ic}^z + C,
\end{split}
\end{equation}
where $S_{ic}^z = X_{ic}-1/2$ is the spin $z$-operator indexed by two variables $(i, c)$; $C=\left[1+K(K-3)/4\right]N+K|E|/4$ is a constant, and $|E|$ denotes the total number of edges. The coefficients $h_i$ are given by
\begin{equation}
 h_i=K+\frac{1}{2}\deg(i)-2
\quad
\text{and}
\quad
J_{ij}
=
\begin{cases}
2 & i=j, \\
1 & i \not = j,
\end{cases}
\end{equation}
where $\deg(i)$ is the number of edges at vertex $i$.

Current quantum annealers, such as the D-Wave machine, are imperfect due to a variety of factors, chief among them is \emph{static disorder} originating in the limited control at the hardware level~\cite{brugger2018quantum,PhysRevX.5.031040,dwave-description}. Therefore, our objective is to investigate what happens to the quantum annealing  when all couplings $J_{ij}$ and magnetic fields $h_i$ are slightly perturbed. To be more specific, we introduce static disorder,
\begin{equation}
\label{eq:disorder}
 h_i \rightarrow  h_i+\delta h_i, 
 \quad 
 J_{ij} \rightarrow J_{ij}+\delta J_{ij}.
\end{equation}
where perturbations $\delta h_i$ and $\delta J_{ij}$ are random variables with flat distributions and symmetric amplitudes, e.g. $\delta J_{ij}~\in~[-W_J,W_J]$ and $\delta h_{i}~\in~[-W_h,W_h]$.

For the sake of simplicity and without any loss of generality we focus in particular on the disorder generator where $\delta J_{ij}=~0$ and moreover (cf. Fig.~\ref{fig:m1m2})
\begin{equation}
h_i\rightarrow \left\{\begin{array}{ll}
h_i+\delta  h_i, & \text{for } h_i+\delta  h_i < \max\{h_i\};\\
\max\{h_i\}, & \text{otherwise}.
\end{array}\right.
\label{eq:dism2}\end{equation}
Such disorder~(\ref{eq:disorder}) mimics to some extent a situation, in which the actual values of interaction strengths at the hardware level differ from the input parameters provided by the programmer operating at the software level.

\begin{figure*}
    \centering
    \includegraphics[width=\textwidth]{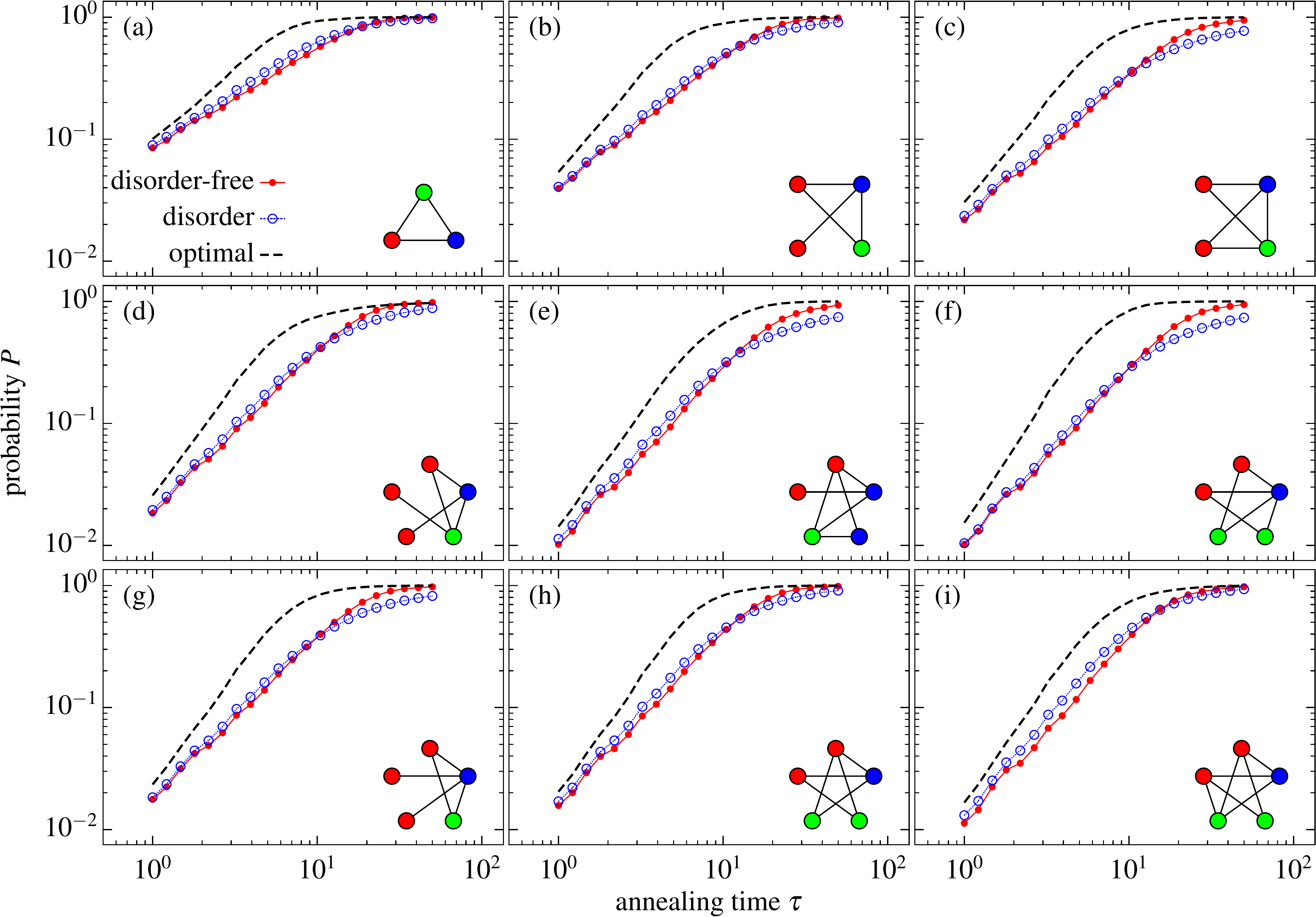}    
 \caption{
 \label{fig:all}
 Probability $P$ to observe the correct final result defined in Eq.~(\ref{eq:projdef}) as a function of the computational/annealing time, $\tau$, for selected problem typologies. Red solid lines correspond to the disorder-free
 case, e.g. $W_h=W_J=0$, whereas blue dashed lines depict results for the disordered case, where magnetic fields are perturbed according to Eq.~\eqref{eq:dism2} with $W_h=1$. Here,  $f(t)=t/\tau$. The corresponding low energy spectra for all three cases are depicted in Fig.~\ref{fig:En}. Black dashed line corresponds to the optimal disorder realization, cf. Fig.~\ref{fig:En}(b).
}
\end{figure*}
\section{Results}
To investigate the dynamics/annealing of the graph coloring problem formulated in Eq.~(\ref{eq:hamcolorspin}), we focus on all non-isomorphic graphs, $G(E,V)$, having $|V|=3,4,5$ vertices and for which the chromatic number $\chi(G)=K>2$. We omit the $K=2$ case as one can reduce its problem Hamiltonian to the antiferromagnetic Ising model.

The quality of a quantum computation/annealing can be measured in various ways~\cite{Santra_2014}. For instance, one may try to count defects~\cite{gardas2018defects},
estimate fluctuations~\cite{gardas2018quantum2}, calculate the fidelity between the final state, $\ket{\psi(\tau)}$, and the true ground state of the problem Hamiltonian~\cite{grass2016quantum}, $\ket{\phi}$, or simply determine the difference between their corresponding energies, 
$\delta E=\bra{\psi(\tau)}\hat{H}\ket{\psi(\tau)}-\bra{\phi}\hat{H}\ket{\phi}$~\cite{kudo2018constrained}. 

In the present work we calculate the probability to observe the correct final result,
\begin{equation}
P  = \sum_{i\in \mathcal{S}} |\langle\psi(\tau)|\phi_i \rangle |^2.\label{eq:projdef}
\end{equation}
Here, $\mathcal{S}$ is a set that labels all possible solutions, $\ket{\phi_i}$, of the disorder-free problem encoded in the Hamiltonian~(\ref{eq:hamcolorspin}).
The final state $\ket{\psi(\tau)}$ is obtained by solving the time dependent Schr\"o{}dinger equation, $i\partial_t\ket{\psi(t)}=\hat{H}(t)\ket{\psi(t)}$, numerically~\cite{NAtTEoCQS,vega}. 
The total Hamiltonian $\hat{H}(t)$ is defined in Eq.~(\ref{eq:AQC}) with the objective Hamiltonian (encoding the graph coloring problem) given by Eq.~(\ref{eq:hamcolorspin}) where all couplings, $J_{ij}$, and biases, $h_i$, are redefined according to Eq.~(\ref{eq:disorder}). 

\emph{A priori}, the disorder amplitudes $W_h$, $W_J$ could be arbitrarily large. However, to ensure that the ground state of the disordered problem matches at least one solution to the disorder-free problem at all, both $W_h$, $W_J$ need to be carefully chosen. For instance, picking $W_J = W_h = 0.5$ guarantees $0.99$ probability of this event to occur (cf. Fig.~\ref{fig:cmap}). For the sake of simplicity, we choose a simple annealing protocol such that $f(t)=t/\tau$. Moreover, we assume without loss of generality that $W_{J}\equiv0$.

\subsection{Disordered energy spectrum}
As depicted in Fig.~\ref{fig:En}, introducing the disorder to the Hamiltonian~(\ref{eq:hamcolorspin}) removes the degeneracy of its ground state. As a result, a solution to the graph coloring problem can be found not only in the degenerate ground state (as in the disorder-free case) but also in low energy spectrum consisting of $M \ll 2^{KN}$ states. In principle, this effect has the potential to increase the overall chances of finding a correct solution, in particular close to the adiabatic limit, e.g. on a time scale $\tau  \sim 1/\Delta$. Here, $\Delta:=E_{i_0}-E_0$ is an effective gap. That is, the difference between the ground state energy $E_0$ and the energy of the first \emph{accessible} state, $E_{i_0}$, which does \emph{not} encode a solution.
\subsection{Disorder-assisted dynamics}
In Fig.~\ref{fig:all} we depict the probability to find the correct answer~(\ref{eq:projdef}) as a function of the annealing time $\tau$ for 
the disordered and disorder-free systems. In the adiabatic limit where $\tau \gg 1/\Delta$, the disorder-free system is more likely to reach the ground state than the disordered one. Nevertheless, introducing disorder into the system does \emph{not} significantly affect the final probability. 

\begin{figure*}
\centering
\includegraphics[width=\textwidth]{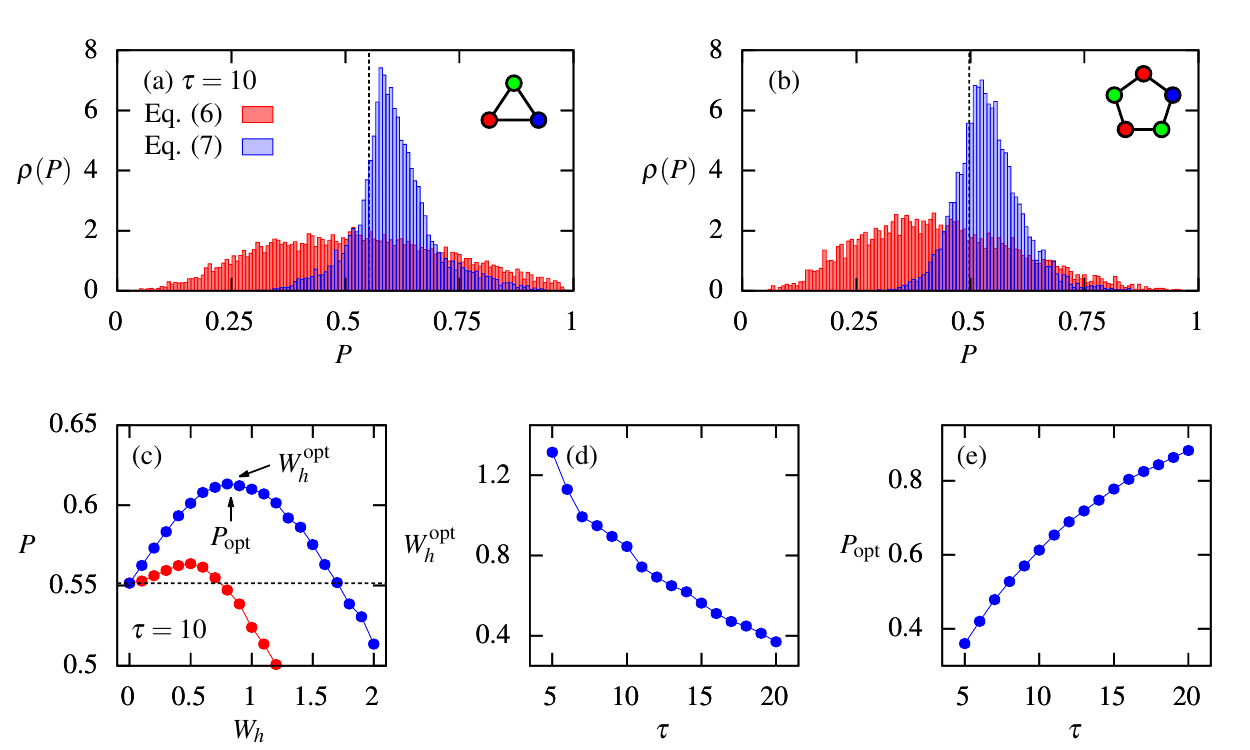}
\caption{(a), (b): Comparison between a generic [Eq.~(\ref{eq:disorder})] and specific [Eq.~(\ref{eq:dism2})] type of disorder for different typologies ($\tau=10$, $W_h=1$ and $W_J=0$).
Here, $\rho(P)$ denotes the density of random variable $P$ defined in Eq.~(\ref{eq:projdef}). 
(c): Influence of the disorder amplitude $W_h$ on probability $P$. 
(d), (e): Annealing time $\tau$ dependence of optimal values $W_h^{\rm{opt}}$ and $P_{\rm{opt}}$.
The result in (a), (c)--(e) and (b) has been obtained for the triangle and pentagon topologies respectively; $W_J=0$.
}
\label{fig:m1m2}
\end{figure*}

On the other hand, for small and moderate $\tau$ we observe that the probability to find the correct solutions is typically larger for the disordered Hamiltonian then in the disorder-free situation. Thus, it is not far-fetched to realize that one can always try to find $\tau_0$ such that $P_{\rm free}(\tau_0)<P_{\rm disorder}(\tau_0)$. This suggests a different strategy to perform computation with noisy near-term quantum annealers. Rather then trying to operate the annealer as adiabatically as possible, one identifies the ``sweet spot'', $\tau_0$, at which the quantum annealer has optimal performance, even better than in the ideal, disorder free case, despite the inevitable noise in the system. For instance, Fig.~\ref{fig:m1m2}(c) indicates a clear maximum. Quite remarkably, we also notice that this is truly a finite-time effect. In Fig.~\ref{fig:m1m2}(d) we plot the optimal value of the noise amplitude as a function of the anneal time. We observe that in the adiabatic limit the disorder-free case is the only ``good'' realization.

However, the impact of the disorder on the success probability is still relatively small. This is illustrated  Fig.~\ref{fig:m1m2}(e). Even at optimal noise strength $P$ is significantly larger for slower processes. Thus, we must ask whether the noise can be modified to make it more ``useful''.
\subsection{Optimizing disorder}
Note that so far we have assumed that noise in the qubit-qubit couplings is uniformly distributed. However, we have also already realized that at intermediate anneal times the presence of noise actually assists the quantum annealer in finding the correct solution. The natural question then is, whether the disorder in the system can be engineered to further enhance this effect---in other words, how to modify the distribution of the noise in our favor. It is then instructive to analyze the energy diagram and dynamics of single realizations of the disordered problem.

To this end, inspect again Fig.~\ref{fig:En}. We observe that in the disorder-free case due to the presence of the degeneracy in the ground state the effective gap $\Delta$ never actually closes, cf. Fig.~\ref{fig:En}(a). The same holds true for ``good'' realizations. Except, that the effective gap opens even wider due to the lack of degeneracy, compare Fig.~\ref{fig:En}(b). On the contrary, for the all the cases we identify as ``bad'', we see some mixture of correct and incorrect solutions that basically behave like impurities causing the effective gap to shrink [cf. Fig.~\ref{fig:En}(c)].
Thus, removing those impurities increases the effective gap which causes the adiabatic threshold to decrease. 

Thus, minimizing the influence of the remaining, ``bad'' realizations may decrease the total time necessary to find a correct solution substantially. This is also demonstrated in Fig.~\ref{fig:all} where the averages dynamics is computed over only those realizations that correlate with corrects solutions. This clearly demonstrates the advantage of disordered dynamics over the ``ideal'', disorder-free situation.
\section{Conclusions}
It is still a commonly accepted creed that noise and disorder in computing hardware have exclusively negative consequences. In the present work, we have shown that this is not always the case, and that static disorder can actually assist quantum annealers in successfully performing their tasks. More specifically, we have studied the graph coloring problem~\cite{kudo2018constrained} on disorder-free and disordered quantum annealers.

On a fundamental level, our results clearly exhibit that moderate noise in the qubit-qubit couplings does not only \emph{not} deter the annealer from finding the correct solution, but also that there are instance where disorder assists the annealer to perform in \emph{finite time}.
A more thorough analysis revealed that in truly adiabatic operation, i.e., for very large anneal times noise is, indeed, detrimental. However, we also found that for short anneal times static disorder can be tuned to significantly enhance the performance of the quantum annealer.

Interestingly, recently a new massively parallel algorithm for simulated annealing has been proposed~\cite{cook2018gpu}. This method contains a non-deterministic element -- lack of synchronization between CUDA threads, which could be (re)interpreted as a source of noise.

On a more practical note, our results may suggest an answer to a conundrum about existing hardware. Systems like the D-Wave machine are known to be subject to electrode noise, which can lead to severe disorder in the on-site fields and qubit couplings. Nevertheless, in particular graph coloring problems have been shown to be solved rather accurately~\cite{dwavedocs,dwavedocs2,dwavedocs3}. A conjecture that can be drawn now is that the D-Wave machine may be operating exactly in such a disorder-assisted regime. 

Of course, further characterization of the D-Wave machine appears necessary to verify our hypothesis. However, if this is indeed the case, then the performance of the machine could be dramatically enhanced by post-selecting the answers on the noise distribution (which will need to be measured independently).
\newpage 
\acknowledgements{
We thank Marcin Mierzejewski and Konrad Jałowiecki for fruitful discussions. This work was supported by the National Science Centre, Poland under projects 2016/23/B/ST3/00647 (AW) and 2016/20/S/ST2/00152 (BG). S.D. acknowledges support from the U.S. National Science Foundation under Grant No. CHE-1648973.
}
\bibliographystyle{apsrev4-1_nature}
\bibliography{lib}

\end{document}